# Charge transfer mechanism and $T_c(x)$ dependence in $Y_{0.8}(Ca)_{0.2}Ba_2Cu_3O_{6+x}$

V. M. Matic, N. Dj. Lazarov and I. M. Bradaric

*Laboratory of Theoretical Physics,
Institute of Nuclear Sciences "Vinca", P.O.Box 522, 11001 Belgrade, Serbia*

**Abstract**

We propose a model for charge transfer mechanism in $Y_{0.8}(Ca)_{0.2}Ba_2Cu_3O_{6+x}$ to count hole doping of $CuO_2$ planes and $x$ dependence of critical transition temperature $T_c$. It is assumed the total number of doped holes in the planes is sum of holes that are introduced through two separate channels: substitution of $Y^{3+}$ by $Ca^{2+}$ and from CuO chains that are longer than a minimal (critical) length $l_{min}$ needed for charge transfer to take place. The $T_c(x)$ dependence is obtained by combining calculated $x$ dependence of doping, $p(x)$, and universal $T_c$ versus $p$ relation. Although calculated $T_c(x)$ dependences for $l_{min}=3$ and $l_{min}=4$ both remarkably correlate to the experimental $T_c(x)$, we argue that the value $l_{min}=4$ gives a reasonable overall agreement.

High temperature superconductivity in cuprates is still a challenging problem in modern solid state physics. The superconducting ground state usually emerges through chemical doping of parent compound which is antiferromagnetic Mott insulator. The fundamental building blocks common to all copper based high temperature superconductors are the $CuO_2$ planes, which have been the subject of a wealth of experimental and theoretical studies. In non-doped parent compounds each Cu atom in the $CuO_2$ plane has one unpaired electron in $3d$ level, so that the net spin of Cu is equal to 1/2. However, unlike in conventional metallic conductors, in which these unpaired electrons can move almost freely from one atomic site to another, the strong Coulomb repulsion between them makes electrons to become localized. On the other hand, given the fact that the energy of the ground state can nevertheless be additionally lowered, although very slightly, when the unpaired electrons make short-lived virtual hops onto the neighboring sites, this hopping appears to be possible only if electrons on the neighboring sites have antiparallel spins (inasmuch as the Pauli exclusion principle prevents electrons to occupy the same quantum state in the same place and at the same time). The breakdown of antiferromagnetic order followed by the onset of superconductivity occurs in these compounds when certain fraction of $3d$ copper electrons, typically ≈5%, is removed from the planes. The number of removed electrons (doped holes) per Cu is conventionally denoted as "doping" $p$, and it is these holes (the missing electrons) that really act as charge carriers, for they are able to move throughout the planes making the material to behave as conductor, or superconductor (at low enough temperatures). Most conveniently the electrons are removed by chemical substitution of metal cations that are sandwiched between the planes, for example, the substitution of $Y^{3+}$ by $Ca^{2+}$ in $Y_{0.8}(Ca)_{0.2}Ba_2Cu_3O_{6+x}$, or $La^{3+}$ by $Sr^{2+}$ in $La_{2-x}(Sr)_xCuO_4$, but the electrons can also be transferred off the planes by introducing oxygen into the material, which orders to form CuO chains in separate chain-layers. These chains are known to act as efficient attractors of electrons (hole suppliers to the $CuO_2$ planes) because of strong tendency of chain oxygen to attract them from other parts of the system.

The hole doping $p$ is one of the most important parameters in high-$T_c$ superconductivity, inasmuch as the doping dependence of various physical quantities has universal character for practically all cuprate families, so that resolving the issue of these dependences has been a major focus of research [1-3]. Typically, when doping rises from $p=0.05$ the critical temperature $T_c$ increases from zero to attain its maximal value $T_{c,max}$ at $p=0.16$. This is accompanied by reduction of pseudo gap energy $E_g$, which is associated with normal state electron correlations above $T_c$, manifested as a sort of depression in electron density of states (also known as "pseudo-gap"). At critical doping level $p=0.19$, the pseudo gap phase eventually vanishes and further increase of doping is characterized by decrease of $T_c$ and termination of superconductivity at $p≈0.27$. As long as the planes are maintained free of additional quasiparticle scatterers (manifested, for example, as a small fraction of Zn atoms that substitute in-plane Cu) the following empirical generic relationship between $T_c$ and $p$ has been found in experiments to hold for a wide class of high $T_c$ cuprates [4,5]

$$T_c(p) = T_{c,max}\left[1 - 82.6(p - 0.16)^2\right]. \qquad (1)$$

The $Y_{0.8}(Ca)_{0.2}Ba_2Cu_3O_{6+x}$ superconductor occupies a specific place among other high-$T_c$ cuprates because the doping is realized through combination of both channels: by substitution of $Y^{3+}$ with $Ca^{2+}$, and by transfer of holes from chains to planes (i.e. transfer of electrons away from planes). Despite the fact that the material has basically the same crystal structure as the parent $YBa_2Cu_3O_{6+x}$ compound, its $T_c(x)$ characteristics reveals apparently different behavior than the famous two-plateaus-like shape of $YBa_2Cu_3O_{6+x}$ [6,7]. The maximal $T_c$ is shifted towards lower oxygen concentrations and, consequently, the highly overdoped regime, extending far beyond optimal doping $p=0.16$, is easier to realize than in the parent material [8]. Furthermore, the $T_c$ versus $x$ dependence in $Y_{0.8}(Ca)_{0.2}Ba_2Cu_3O_{6+x}$ shows a plateau feature in oxygen composition interval $0<x<0.35$ [8,9], which appears to be drastically shifted comparing to the one of the $T_c(x)$ of the $YBa_2Cu_3O_{6+x}$ compound (incidentally, it is almost exactly where the $T_c$ of the latter compound falls to zero). In the light of the fact that the long standing controversy about the origin of the plateaus at 60K and 90K in $T_c(x)$ of $YBa_2Cu_3O_{6+x}$ has finally been resolved in recent publications [22,26], it poses itself as a challenging further step to address the issue of strikingly different $T_c(x)$ in homologous $Y_{0.8}(Ca)_{0.2}Ba_2Cu_3O_{6+x}$ by applying basically the same microscopic mechanism, although suitably modified to include doping contribution that comes from introduced Ca. Besides of altered $T_c(x)$ characteristics, it would be also interesting to know whether the effective charge redistribution, initiated by Ca inserted at Y sites, is confined only to Y layer and to two $CuO_2$ layers (between which it is sandwiched), or it extends further away, up to the Ba layer and the chain plane, affecting the net chains efficiency to conduct hole transfer to the planes. Addressing these two issues is the main objective of this study.

    We propose here a theoretical model to count number of holes generated in $CuO_2$ planes of $Y_{0.8}(Ca)_{0.2}Ba_2Cu_3O_{6+x}$. Calculated $p(x)$ is combined with universal relation (1) to yield $T_c(x)$ that remarkably agrees with experimental result. The proposed model assumes that the net hole concentration in the planes is sum of two terms: the term that includes holes introduced through the replacement of $Y^{3+}$ by $Ca^{2+}$ and the term referring to the holes that originate from CuO chains. It is also assumed that only chains whose length is greater than, or equal to, a certain minimal length $l_{min}$ (needed to trigger chain-to-plane charge transfer), can effectively supply holes to the $CuO_2$ layers. Our results point to the conclusion that the most likely values for $l_{min}$ are 3, or 4 (three, or four oxygen atoms in a chain). We also show that the averaged ability of a chain hole to attract an electron from $CuO_2$ planes is diminished by $\approx 20\%$ compared with the parent (Ca free) material.

    It is widely accepted opinion that copper in basal (chain) planes can exist either as $Cu^{1+}$, which occurs when it is not linked to in-plane oxygen (but only to two apical O(4) anions), or as $Cu^{2+}$, which is the case when it is bonded to two, or to one, in-plane oxygen (the 4-fold, or the 3-fold, coordinated Cu, respectively). Since oxygen has very strong tendency to accept two electrons from its surroundings and, consequently, to stabilize $O^{2-}$ valence state, it follows that CuO chain that contains $l$ oxygen anions, i.e. the chain of length $l$, has capacity of attracting $l-1$ electrons from other parts of the system. These missing electrons along the chain are usually referred to as "holes", so that the chain of length $l$ is considered as to have created $l-1$ holes that are then capable of attracting electrons from other parts of the system, presumably from two $CuO_2$ layers (since each transferred electron leaves a hole in the layers the net effect is that a hole appears to be transferred from the chain to the layers). It seems fairly reasonable to expect that in the system that contains longer chains and, consequently, more chain-holes per basal plane Cu, there would be a greater probability for a hole to hop onto the layers and therefore to

produce higher hole concentrations in them. Indeed, such a point of view has unambiguously been supported in experiments on photoinduced persistent superconductivity [10-13] and room temperature aging of samples that had previously been rapidly cooled from high temperatures [14,15], which both have inferred that longer chains, emerging through the corresponding processes of oxygen reordering (without changing the oxygen content $x$), produce larger hole transfer and therefore higher $T_c$ values. Furthermore, the chains are believed that can efficiently supply holes to the planes only if they are long enough [6]. Such an idea has spontaneously emerged through attempts to provide a reasonably acceptable explanation for the existence of 60K plateau in YBa$_2$Cu$_3$O$_{6+x}$ compound, that occurs in so-called ortho-II phase (at $x \geq 0.5$). In this phase long CuO chains alternate along $a$-axes with long sequences of vacant oxygen sites (occupied sites are usually referred to as "$\alpha_1$", and vacant sites as "$\alpha_2$"), so that when $x$ increases beyond $x=0.5$ additional oxygen atoms occupy $\alpha_2$ sites in a random fashion. As a consequence, at $x>0.5$ the $\alpha_2$ sublattice hosts mainly isolated oxygen and, to a lesser degree, short CuO chains, implying that the number of holes induced in the CuO$_2$ planes remains fairly constant until $\alpha_2$ chains become long enough to trigger additional charge transfer [6,14,15]. Such a scenario is therefore presupposed upon two underlying premises: a) the 60K plateau in the parent YBa$_2$Cu$_3$O$_{6+x}$ system is connected with constant doping level ($p(x) \approx const=0.094$ [26]), in the regime of ortho-II phase, and b) there exists certain minimal chain length $l_{min}$ so that only chains of length $l \geq l_{min}$ can induce holes in the planes, in contrast to those with $l<l_{min}$. Although these two assumptions have recently been successfully used to account for both 60K and 90K plateaus in YBa$_2$Cu$_3$O$_{6+x}$ [22,26], we are to exploit here only the latter one in the task of resolving the issue of $T_c(x)$ dependence in Y$_{0.8}$(Ca)$_{0.2}$Ba$_2$Cu$_3$O$_{6+x}$.

In principle, a chain of length $l$ can be created in a number of different ways, as for example, by merging two or more shorter chains, or even by adding oxygen atoms one by one. Whatever the way the chain has been constructed it is reasonable to assume that the final state of chain electronic subsystem ought to be one and the same. Since this state is supposed to be mirroring the net efficiency of chain to attract electrons, i.e. to conduct transfer of holes to the planes, it appears that the most advisable way for following the development of chain's hole transfer efficiency as chain length $l$ gradually increases, is to think about the chain as it has been formed by adding oxygen one by one. Thus, when $l$ increases starting from $l=1$ (isolated oxygen) the minimal chain length concept presupposes that no charge transfer takes place unless $l=l_{min}$. This implies that creation of first $l_{min}-2$ chain-holes is not accompanied by transfer of any charge to (from) the chains. At $l=l_{min}$ the charge transfer becomes initiated and it develops further as chain length continues to increase to its final value, during which process the remaining $l-l_{min}+1$ chain-holes are created. We call these holes *the active holes* because their creation develops simultaneously with the ongoing transfer of electrons (holes). Therefore, in each long chain ($l \geq l_{min}$) we distinguish two types of chain-holes: the first $l_{min}-2$ holes we call *the passive holes*, for their creation does not coincide with any charge transfer, and the remaining $l-l_{min}+1$ active holes that in principle can attract electrons.

However, one should not be misled into thinking that each active chain-hole is to succeed in capturing an electron from the CuO$_2$ planes, since the available experimental data [8,9] in a persuasive way indicate that it is far below one half of them that will eventually manage to accomplish this task. Consider the situation at $x \approx 1$ (ortho-I stoichiometry) where long chains are known to prevail all over the material in $Y_{0.8}(Ca)_{0.2}Ba_2Cu_3O_{6+x}$ system (and also in the parent $YBa_2Cu_3O_{6+x}$). The average chain length $l_{av}=2x/n$ is very large here ($n$ denotes the fraction of 3-fold coordinated Cu in chain plane), which is conventionally formulated in the way that $n\to 0$, as $x\to 1$. Given the fact that concentration of passive holes $h_p$, defined as their number per basal plane Cu (or, equivalently, per Cu in $CuO_2$ plane), cannot exceed $(n/2)(l_{min}-2)$ [26], it therefore behaves as vanishing quantity as $x\to 1$, regardless of what the value of the parameter $l_{min}$ might be equal to. Taking into account that concentrations of active and passive chain holes (denoted by $h$ and $h_p$, respectively; both $h$ and $h_p$ being defined as their number per Cu) are connected with $x$ by an obvious relation $h+h_p+(n/2)=x$, it follows that at $x\approx 1$ practically each oxygen atom has introduced one active hole, so that the concentration $h$ is also equal to 1. Inasmuch as each chain plane supplies holes to two $CuO_2$ layers, the net chain contribution to doping would be very close to 0.5 at $x\approx 1$, if it were that each active hole succeeded to attract one electron. Experiments, however, clearly oppose such a scenario, since even the combined contribution of both chains and Ca has been found to be lesser than 50% of anticipated 0.5 (at $x\approx 1$). To see that, equation (1) is to be applied to reported experimental results on $T_c(x)$ of $Y_{0.8}(Ca)_{0.2}Ba_2Cu_3O_{6+x}$ [8,9] to obtain that at $x\approx 1$ the concentration of holes induced in the $CuO_2$ layers is ranking around $\approx 0.24$ (see Figure 1). Therefore, the doping at $x\approx 1$, $p(x\approx 1)\approx 0.24$, is far less than 0.5, pointing to the conclusion that in long chains it is only a fraction, surely less than one half, of chain created holes that will succeed in capturing electrons from the planes. In order to estimate this fraction one must first extract the Ca contribution from $p(x\approx 1)\approx 0.24$. To do that, the generic relation (1) is to be applied again, but this time on reported $T_c(x)$ [8,9] in the regime $x\approx 0$ (Figure 1), in which the chain contribution to charge transfer is expected to have faded away, and the total doping, and, consequently, the plateau feature in $T_c(x)$, to be stemming only from Ca. Doing this way, one obtains $p(x\approx 0)\approx 0.078$, which is a rather surprising result since one would expect $p(x\approx 0)\approx b/2=0.1$ ($b=0.2$ in $Y_{1-b}(Ca)_bBa_2Cu_3O_{6+x}$). This indicates that in fact not all holes introduced by Ca are transferred to the $CuO_2$ planes, but only $\approx 78\%$ of them. At this stage we do not have a definite idea about where the remaining $\approx 22\%$ of Ca holes should be assigned to (the most likely, they are to stay localized close to Ca), but this $\approx 22$ percent departure from the expected value seems to be a rather systematic result because it equally applies to the substitution level $b=0.1$ (i.e. in $Y_{0.9}(Ca)_{0.1}Ba_2Cu_3O_{6+x}$), as it can be easily verified from corresponding data on experimentally measured $p(x)$ [4] (it should be noted that unlike in $Y_{0.8}(Ca)_{0.2}Ba_2Cu_3O_{6+x}$, where $p(x)$ was not directly measured but we obtained it here from measured $T_c(x)$ of references [8,9] by applying relation (1), the $p(x)$ was directly determined in $Y_{0.9}(Ca)_{0.1}Ba_2Cu_3O_{6+x}$ by measuring interatomic distances and calculating bond-valence sums [4]). Therefore, at $x\approx 1$, the doping appears to be composed of two contributions, $p(x\approx 1)=(b_{eff}/2)+\chi/2$, where $b_{eff}$ stands for $0.78b=0.156$ and $\chi$ denotes the fraction of chain holes that have been transferred to the $CuO_2$ bilayer. Introduced in this way, the quantity $\chi$ measures the effectiveness of an active hole to attract an $3d$ electron from the planes. From the above analysis it follows that $\chi$ should be ranking around 33%.

Although estimated value of the active hole efficiency, $\chi\approx 0.33$, has now been extracted from the state in which long chains prevail ($x\approx 1$), we introduce here an additional assumption that $\chi$ has the same value in chains of all possible lengths, i.e. from $l=l_{min}$ to $l=\infty$, so that the number of transferred holes from a chain of length $l\geq l_{min}$ appears to be equal to $\chi(l-l_{min}+1)$ [22,26]. If $f(l)$ denotes fraction of chains that have the same length $l$, the total number of transferred holes $N$ would then be equal to $(\chi n/2)N_{Cu}\sum_{l=l_{min}}^{\infty}(l-l_{min}+1)f(l)$, where $N_{Cu}$ stands for the number of Cu in basal plane. In the case of ortho-II structural phase, whose existence in the $Y_{0.8}(Ca)_{0.2}Ba_2Cu_3O_{6+x}$ system has been experimentally confirmed [16], the oxygen sites split into two interlacing sublattices commonly known as $\alpha_1$ and $\alpha_2$ (as mentioned above). These sublattices are characterized by different rates of chain breaking $n_1 \neq n_2$ ($n=(n_1+n_2)/2$), different oxygen occupancies $x_1 \neq x_2$ ($x=(x_1+x_2)/2$) and, consequently, different average chain lengths $l_{av,\alpha 1}=2x_1/n_1$ and $l_{av,\alpha 2}=2x_2/n_2$. The doping $p=N/2N_{Cu}$ is then equal to

$$p = \frac{b_{eff}}{2} + \frac{\chi}{8}\left[n_1 \sum_{l=l_{min}}^{\infty}(l-l_{min}+1)f_{\alpha 1}(l) + n_2 \sum_{l=l_{min}}^{\infty}(l-l_{min}+1)f_{\alpha 2}(l)\right], \quad (2)$$

where $f_{\alpha 1}(l)$ and $f_{\alpha 2}(l)$ denote length distributions of CuO chains on corresponding oxygen sublattices. In the above equation the first term, $b_{eff}/2$, describes the contribution to doping that comes from Ca substitute, while the second term refers to the chain contribution $p_{ch}$ (that is associated with the charge transfer from chains to planes).

We used two dimensional asymmetric next nearest neighbor Ising (ASYNNNI) model to describe thermodynamics of CuO chain formation because this model has long been known that stabilizes both major orthorombic phases, OI and OII, as its ground state [17]. The length distributions of CuO chains are known to obey the following geometric-like behavior: $f_{\alpha i}(l)=\omega_i(1-\omega_i)^{l-1}$, $i=1,2$, where $\omega_i$ denotes the inverse of average chain length, $(l_{av,\alpha i})^{-1}$, on corresponding oxygen sublattice [18]. Such behavior of length distributions ensures a rapid convergence of summations in equation (2). Although theoretical study has shown that a certain departure from geometric-like behavior of $f(l)$s can be expected in narrow interval $\Delta x \approx 0.07$ around critical point of the OI-to-OII phase transition, due to increased fluctuations of energy of the ASYNNNI model [18], our extended analysis has inferred that such departures are nevertheless compensated by sums in (2) [26], so that the doping can be expressed in the following integrated form

$$p = \frac{b_{eff}}{2} + \frac{\chi}{4}\left[x_1\left(1-(l_{av,\alpha 1})^{-1}\right)^{l_{min}-1} + x_2\left(1-(l_{av,\alpha 2})^{-1}\right)^{l_{min}-1}\right]. \quad (3)$$

We calculated doping (3) using estimated values of $b_{eff} \approx 0.156$ and $\chi \approx 0.33$. In fact we treated both of these parameters as quantities that should be varied, although only slightly, around their expected values in order to achieve the best fitting between calculated and experimental $T_c(x)$'s. For given $x$ and $T$, the quantities $x_1$, $x_2$, $n_1$, and $n_2$ in (3) were determined by use of the cluster variation method (CVM) in approximation of six 5/4 point basic clusters, for O-O interaction parameters of the ASYNNNI model ($V_1>0$ (nearest neighbor), $V_2<0$ (copper mediated next nearest neighbor), and $V_3>0$ (repulsive Coulomb next nearest neighbor)), as obtained by linear muffin tin orbital method (LMTO) for the case of $YBa_2Cu_3O_{6+x}$ compound [19]. Although the LMTO values of the interaction parameters were sometimes questioned even in the case of the $YBa_2Cu_3O_{6+x}$ system [20], we nevertheless used them here merely as a probe to check applicability of the model expressed by (2) and (3). It should be noted, however, that the question of magnitudes of pairwise O-O interactions in $Y_{0.8}(Ca)_{0.2}Ba_2Cu_3O_{6+x}$ compound is still open particularly in view of the fact that there has been no detailed experimental study on phase dislocations in $(x,T)$ space for this system thus far. Accordingly, it is difficult to estimate interactions from specific features of the phase diagram, for they are in fact not known exactly (e.g. the region of stability of ortho-II phase, and, particularly, the temperature $T_{OII}$ that corresponds to the top of the ortho-II phase in $(x,T)$ space).

Beside the interaction constants $V_i$, $i=1,2,3$, the value of the parameter $l_{min}$ is, according to (3), also required to determine $x$-dependence of doping, $p(x)$, and so as the value of the reduced temperature parameter, $\tau=k_BT/V_1$, that should be referred to room temperature. As far as $l_{min}$ is concerned, it should be noted that in our recent study [26] it has been shown that at any $\tau=const$ there exists a well defined value of $l_{min}$, which we denoted by $l_{opt}(\tau)$ (the so-called *optimal* chain length), for which $p_{ch}(x)$ remains constant as $x$ increases beyond $x\approx0.5$ over the region of ortho-II phase. From Figure 1 of reference [26] it can be seen that, at a given $\tau=const$, if $l_{min}< l_{opt}(\tau)$ then $p_{ch}(x)$ is monotonically increasing function over the ortho-II phase region (and also in the whole interval $0<x<1$). However, if $l_{min}>l_{opt}(\tau)$, then $p_{ch}(x)$, and, consequently, the concentration of active chain holes, both have a maximum at $x\approx0.5$, that is followed by a decline when $x$ increases beyond 0.5, to turn to be rising again after $x$ enters into the regime of ortho-I phase [22,26]. Such a peaky behavior of $p(x)$ at $\tau=const$ is clearly not what generates the $T_c(x)$ in $Y_{0.8}(Ca)_{0.2}Ba_2Cu_3O_{6+x}$ system, but it is the monotonically increasing $p(x)$ (and $p_{ch}(x)$), as it can be straightforwardly verified by applying equation (1) on experimentally obtained $T_c(x)$ [8,9] (the so obtained $p(x)$ is shown by open squares in Figure 1). This points to the conclusion that in case of $Y_{0.8}(Ca)_{0.2}Ba_2Cu_3O_{6+x}$ system the quantities $l_{min}$ and $\tau_{RT}$ ($\tau_{RT}$ stands for the value of $\tau$ that refers to room temperature) should be ascribed such values to secure the relation $l_{min}<l_{opt}(\tau_{RT})$. In order to estimate $\tau_{RT}$ for the $Y_{0.8}(Ca)_{0.2}Ba_2Cu_3O_{6+x}$ system it is worthwhile to recall that in the case of $YBa_2Cu_3O_{6+x}$ the top of ortho-II phase (above $x\approx0.5$) is very well known to occur at $\approx125\div140°C$ [28], which places $\tau_{RT}\approx0.45$ to be a fairly reliable estimation [26], since the LMTO nearest neighbor O-O interaction $V_1$, that ranges around $\approx6.7\div6.9mRyd$ [19.20], fixes scaling between $T$ and $\tau$ in a way that $\Delta\tau\approx0.1$ corresponds to $\Delta T\approx100K$, and also because the theoretically obtained phase diagram sets up the top of ortho-II phase at $\tau\approx0.58$ [21]. In $Y_{0.8}(Ca)_{0.2}Ba_2Cu_3O_{6+x}$ system a precise location of the top of ortho-II phase is not really known, although a clear ortho-II signal has unambiguously been observed in experiments [16], and, furthermore, there have been no studies reported thus far on how the presence of Ca would alter the effective O-O interactions in chain plane. Therefore, it is difficult to make a reliable estimation for $\tau_{RT}$ and, thus, to set up the upper limit for $l_{min}$ through $l_{opt}(\tau_{RT})$. In situation like this, we relied on the following strategy: Firstly, we retained the LMTO values of in-plane O-O interactions that were successfully used in $YBa_2Cu_3O_{6+x}$ [19] (although we were well aware that the effective $V_1$, $V_2$, and $V_3$ interactions might indeed be of different values in $Y_{0.8}(Ca)_{0.2}Ba_2Cu_3O_{6+x}$), because we primarily wanted to examine applicability of the model (3) when combined with the ASYNNNI model of in-plane oxygen ordering, and, secondly, given the fact that $l_{opt}(\tau)$ rapidly increases with $\tau$ lowering (as shown in Figure 1 of reference [26] - the results of our the most recent analyses show that $l_{opt}(\tau)$ scales with $\approx\exp(2|V_2|/k_BT)$), we attached to $\tau_{RT}$ a lower value than in $YBa_2Cu_3O_{6+x}$ system in order to open a way for a broader range of $l_{min}$ values to produce monotonically increasing $p_{ch}(x)$ and $p(x)=(b_{eff}/2)+p_{ch}(x)$. We thus arrived at $\tau_{RT}=0.38$ that might be adopted as a relatively acceptable estimation for $\tau_{RT}$ as it opens a way for $l_{min}=3, 4$, or $5$ to be producing a monotonic behavior of $p_{ch}(x)$, since at $\tau=0.38$ the $l_{opt}(\tau)$ falls at some point between 6 and 7 (as shown in Figure 1 of reference [26]). It should be noted, however, that had the $\tau_{RT}$ been ascribed a lower value than 0.38, it would have obviously broadened the range of possible values for $l_{min}$, but we did not want to go too low with $\tau_{RT}$, for that would imply as if the ortho-II phase is more pronounced in $Y_{0.8}(Ca)_{0.2}Ba_2Cu_3O_{6+x}$ than in $YBa_2Cu_3O_{6+x}$, and yet such a development has not been grounded, at least in a persuasive way, upon available experimental data [16] (although the ortho-II phase has indeed been clearly detected in $Y_{0.8}(Ca)_{0.2}Ba_2Cu_3O_{6+x}$, as mentioned previously). Furthermore, taking into account that $l_{min}=4$, or, alternatively, $l_{min}=5$, have recently been shown to successfully account for both 60K and 90K plateaus in the $YBa_2Cu_3O_{6+x}$ system [22,26], we were of opinion that introduction of ≈20% of Ca ions would not affect that drastically the chain's ability to trigger the charge transfer process, i.e. to make the value of the $l_{min}$ parameter significantly altered from that of the parent compound. Therefore, the estimated $\tau_{RT}=0.38$ makes $l_{min}=3$, 4, or 5 appearing as obvious candidates to be lying at the root of $T_c(x)$ characteristics in $Y_{0.8}(Ca)_{0.2}Ba_2Cu_3O_{6+x}$ system.

Calculated $p_{ch}(x)$ dependences at $\tau=0.38=const$, for $l_{min}=3$, 4, 5 and 6, are shown in Figure 1 by solid lines. The $p_{ch}(x)$ for $l_{min}=3$ is shown at the bottom (this case corresponds to $b_{eff}=0$ in (3)), while the $p_{ch}(x)$'s for $l_{min}=4$, 5 and 6 are shifted upwards for $b_{eff}/2=0.12$, 0.18 and 0.24, respectively, towards the top of the Figure in order to avoid overlapping, so that their flow can be followed transparently. In addition, the total doping, $p(x)=(b_{eff}/2)+p_{ch}(x)$ for $l_{min}=3$ and $b_{eff}=0.154$ is also shown in the middle section (the curve that overlaps with a sequence of open squares which stands for the $p(x)$ extracted from experimental $T_c(x)$ [9]). The three calculated $x$-dependences of the total doping, $p(x)=(b_{eff}/2)+p_{ch}(x)$, for $l_{min}=3$, 4, and 5, and for $b_{eff}=0.154$ ($\chi=0.336$), 0.155 ($\chi=0.334$), and 0.156 ($\chi=0.334$), respectively (the latter two not shown in Figure 1 for correct values of $b_{eff}$) were then combined with the generic $T_c$ versus $p$ relation (1) (for $T_{c,max}=85.5K$) to obtain three $T_c(x)$ dependences that are shown in Figures 2a, 2b, and 2c, respectively. For obtained $h_a(x)$ dependences ($p_{ch}(x)=(\chi/2)h_a(x)$ and $p(x)=(b_{eff}/2)+p_{ch}(x)$) the values of parameters $\chi$ and $b_{eff}$ were finally adjusted by varying them around their expected values (0.33 and 0.156, respectively) in order to achieve the best correlation of calculated $T_c(x)$'s with the experimental one [9]. For three calculated $T_c(x)$ dependences it was obtained that $\chi$ varies between 0.334 and 0.336, and $b_{eff}$ varies between 0.154 and 0.156, which suggests that both $\chi$ and $b_{eff}$ have in fact very well defined values that are practically indistinguishable from those that were estimated beforehand (it should be noted that, if the parameter values were altered even to a slightly higher degree from the quoted ones, the correlation between calculated and experimental $T_c(x)$'s would be lost in a visually apparent fashion - either at the overdoped, or at the underdoped side of $T_{c,max}$).

From Figure 1 it can be seen that the calculated $p_{ch}(x)$'s and $p(x)$ reveal a kink feature, manifested as a change of slope at $x≈0.5$ (the ortho-II stoichiometry), that becomes more pronounced as $l_{min}$ increases. The kink progressively displays a tendency to develop into a horizontal section at $x>0.5$ as $l_{min}$ approaches $l_{min}=6$, which is not surprising given the fact that at $\tau=0.38$ the $l_{opt}(\tau)$ falls at some point between 6 and 7 [26]. Notably, the $p(x)$ extracted form experimental $T_c(x)$ [9] (open squares) also undergoes a change of slope around $x≈0.5$, although not that pronounced as calculated ones, but it nevertheless suggests that the model (3) successfully accounts for this important characteristics of $p(x)$ dependence in $Y_{0.8}(Ca)_{0.2}Ba_2Cu_3O_{6+x}$. In particular, the $p(x)$ for $l_{min}=3$ displays a somewhat less pronounced change of slope that correlates fairly well with the experimental one (a smaller change of slope is quite understandable in view of the fact that $l_{min}=3$ lies further away from $l_{opt}(\tau)=6(7)$ than, for example, $l_{min}=5$). Regarding the above facts, it is not surprising that at first glance the calculated $T_c(x)$'s remarkably fit to experimental values particularly for $l_{min}=3$ and to a lesser degree for $l_{min}=4$ (Figures 2a and 2b), while for $l_{min}=5$ the disagreement looks as being a bit more pronounced (Figure 2c). Although all of the three calculated curves follow fairly well the general flow of experimental $T_c(x)$, the $l_{min}=5$ curve demonstrates a wide flat section that extends approximately between $x \approx 0.5$ and $x \approx 0.7$ instead of to display a clear maximum at $x \approx 0.55$ (or at a value that is slightly higher than that, but not to exceed $x \approx 0.6$). This is obviously due to the fact that the corresponding $p_{ch}(x)$ reveals an expressive tendency towards being horizontal at $x>0.5$ (Figure 1). On the other hand, unlike to the $l_{min}=5$ case, the $p_{ch}(x)$ for $l_{min}=3$ intersects the optimal doping level $p=0.16$ by a larger angle, so that the corresponding $T_c(x)$ shows a maximum that is more distinctively featured and, consequently, correlates better to the experimental result [9] around $T_{c,max}$ (Figure 2a). To get a clearer idea about the overall agreement between the calculated and experimental $T_c(x)$'s we used, as a rough estimate, the quantity $\sigma = \sqrt{(1/M)\sum_{i=1}^{M}(T_c(x_i) - T_{c,\exp}(x_i))^2}$, where $M=20$ stands for the number of experimental points in Figures 2a-c. Although the obtained values $\sigma_1=5.56$, $\sigma_2=4.25$, $\sigma_3=4.48$, for $l_{min}=3$, 4, 5, respectively, point to the conclusion that $l_{min}=4$ gives the best overall agreement, and, consequently, $l_{min}=3$ the worst, we are of opinion that the latter value should not be definitely discarded for the following reasons: All three values of $\sigma$ are of the same order of magnitude, but for $l_{min}=4$ and 5 there are 4-5 points which practically contribute zero terms in the summation for $\sigma$, whereas such points are virtually absent in the $l_{min}=3$ case (as it is evident from Figures 2a,b, when compared with Figure 2c). On the other hand, a closer inspection of the data shows that a slightly greater $\sigma$ for $l_{min}=3$ is mainly due to a rather systematic overshooting of obtained $T_c(x)$ on the overdoped side, which is nevertheless not all that large. Speaking in general terms, in heading towards better agreement one typically wants to hit with the theoretically obtained curve as more experimental points as possible, but attaining a greater fraction of ideally reproduced points may sometimes be achieved at the cost of more pronounced disagreement at some other points. Such typical situation is what we have here for $l_{min}=5$ (and partially for $l_{min}=4$), whose $T_c(x)$ considerably departs from experimental values around $x \approx 0.5$, in contrast to $T_c(x)$ of $l_{min}=3$ that does not display such a deviation (Figure 2c). Instead, the latter curve features a rather round shape at its top, reproducing fairly accurately the behavior of the experimental $T_c(x)$ around $T_{c,max}$ (being virtually free of any artificial wide flatness whatsoever). Given the above facts, we are of opinion that $l_{min}=3$ and $l_{min}=4$ emerge as the most probable candidates for the $l_{min}$ parameter to be utilized in (2) and (3), although the value $l_{min}=4$ gives better overall agreement.

In conclusion, we have shown that the model (3) reproduces fairly well the $p(x)$ and $T_c(x)$ characteristics in $Y_{0.8}(Ca)_{0.2}Ba_2Cu_3O_{6+x}$ compound. The best fitting is achieved here for $l_{min}=3$ and 4, although it should be noted that these conclusions are to a great deal dictated by our estimation on $\tau_{RT}$. If a lower value than 0.38 were assigned to $\tau_{RT}$, and consequently $l_{opt}(\tau)$ attained a value beyond 6 or 7, that would open a way for higher values of $l_{min}$ to produce the $p(x)$ with the change of slope at $x \approx 0.5$ that would not be that large as for $l_{min}=5$ at $\tau=0.38$ (as shown in Figure 1). For example, as it has been mentioned above, at $\tau=0.30$ the corresponding $l_{opt}(\tau)$ falls at some point between 11 and 12 [26], which would allow even $l_{min}=7$ and $l_{min}=8$ to emerge as relatively promising candidates to produce a smaller change of slope in $p(x)$. However, the estimated $\tau_{RT}=0.30$ would mean that the top of ortho-II phase is much higher in $Y_{0.8}(Ca)_{0.2}Ba_2Cu_3O_{6+x}$ ($\approx 280°C$) than in $YBa_2Cu_3O_{6+x}$ ($\approx 125°C$ [28]), which would in turn imply that the ortho-II signal would have to be stronger in $Y_{0.8}(Ca)_{0.2}Ba_2Cu_3O_{6+x}$ than in $YBa_2Cu_3O_{6+x}$ (such a statement, as far as we know, has never been clearly announced in the literature). Besides of that, as our extended analysis indicates, if the value of $l_{min}$ were taken to be too large it would additionally reduce doping $p(x)$ (particularly in the region of $x$ approximately between 0.7 and 1), for it means that more terms would be taken away from sums in (2). As a consequence, at $x>0.7$ the $p(x)$ would assume a more concave form additionally sharpening the angle by which it approaches the $x=1$ axes (in closing to its $l_{min}$-independent value, $p(x\approx 1)=\chi/2$), which would in turn make it practically impossible to obtain coordination between the calculated and experimental $T_c(x)$'s in highly overdoped regime (moreover, it would also shift the optimal doping level $p=0.16$, and, consequently, the maximal $T_c$, to higher values of $x$). Therefore, the wholeness of our results suggests that the most likely values for the $l_{min}$ parameter in $Y_{0.8}(Ca)_{0.2}Ba_2Cu_3O_{6+x}$ system are $l_{min}=3$, or $l_{min}=4$. It should be underlined, however, that even $l_{min}=5$ could be taken into consideration, if $\tau_{RT}$ were to be assigned a lower value than 0.38. Although some theoretical studies suggest it might be $l_{min}=3$ [27], it should be pointed out that resolving this issue decisively, i.e. whether $l_{min}=3$, 4, or even 5, is to become possible only after a more detailed study on structural phase diagram of $Y_{0.8}(Ca)_{0.2}Ba_2Cu_3O_{6+x}$ will have been made (in particular, determining the top of the ortho-II phase).

The very fact that our estimation for $l_{min}$ in $Y_{0.8}(Ca)_{0.2}Ba_2Cu_3O_{6+x}$ converges either to $l_{min}=3$, or to $l_{min}=4$, might rise some controversy in the light of our recent findings for $YBa_2Cu_3O_{6+x}$ superconductor that the two-plateaus phenomenon of its $T_c(x)$ is associated with $l_{min}$ being equal either to 4, or to 5 [26]. Given that the parameter $\chi$, expressing the averaged ability of a chain-hole to attract an electron form $CuO_2$ planes, was found in $YBa_2Cu_3O_{6+x}$ to be lying at around $\approx 41\%$, we made a comment in reference [26] that the latter value of $l_{min}$ is, possibly, to a slightly higher degree to be believed in, since both $l_{min}=5$ and $\chi\approx 2/5$ appear to coherently suggest that three chain-holes are not sufficient to effectuate a hop of $3d$ electron (from the planes to the chain), but that five of them suffice to accomplish transfer of two electrons (one from the plane above, and the other one from the plane below the chain). Such a reasoning would lead to an idea that $l_{min}$ in $Y_{0.8}(Ca)_{0.2}Ba_2Cu_3O_{6+x}$ should be equal to 4, since both $l_{min}=4$ and $\chi\approx 1/3$ infer that two holes cannot attract an electron, but that three of them are able to attract one. However, as our preliminary results indicate, in homologous $Y_{0.9}(Ca)_{0.1}Ba_2Cu_3O_{6+x}$ ($b=0.1$) system the parameter $\chi$ ranges around 36%, which is the value that makes such reasoning rather complicated, assuming, of course, that $l_{min}$ is not to be ascribed a value that would be too large, e.g. extending far beyond 4, or 5. Therefore, even though $\chi$ can be in principle estimated very accurately in $Y_{1-b}(Ca)_bBa_2Cu_3O_{6+x}$ systems (in fact, $\chi$ is to be taken as $b$-dependent quantity, $\chi=\chi(b)$), it alone cannot be used as a reliable ground for resolutely clearing up of what is the concrete value that the parameter $l_{min}$ should be equal to. On the other hand, in view of the existing shortage of an additional decisive criteria to convincingly determine $l_{min}$ (apart from the basic one, which states that the correct $l_{min}$ should produce $T_c(x)$ that agrees with experiment), it seems reasonable to pose a question of whether is it in principle justifiable to expect $l_{min}$ to have *a universal* value, that would be one and the same in all such otherwise different $Y_{1-b}(Ca)_bBa_2Cu_3O_{6+x}$ systems (e.g. $b$=0, 0.1, or 0.2), or, perhaps, $l_{min}$ should be conceived as a quantity whose value is to be affected by the concentration $b$ of Ca ions? Before answering to this question, it is worthwhile to make a note that, generally, some chain characteristics are normally expected to be highly influenced by physical conditions in the chain environment, while other can be anticipated to be weakly influenced, or practically independent (the latter should be regarded as being related to authentically internal chain degrees of freedom). As far as the parameters $l_{min}$ and χ are concerned, it is evident that they both determine various aspects of interaction between the chain and the rest of the system, since they characterize the chain's ability to conduct transfer of charge to (from) other system units. Thus, the length dependent ability of a given chain, determining whether the chain will participate in the charge transfer process, or not (as expressed by the parameter $l_{min}$), is very likely to be affected by the net charge redistribution that takes place along *c*-axes when Ca ions are introduced onto the Y sites (the Y layer is embedded between two $CuO_2$ planes). The redistributed charge significantly alters the charge balance between the chain layer and $CuO_2$ planes, as manifested by different rate of charge transfer between them and different $T_c(x)$ characteristics (in comparison with the parent $YBa_2Cu_3O_{6+x}$ system). Thus, the charge deployed in the close vicinity of chains appears to be rearranged and it is therefore not to be expected in advance that $l_{min}$ should have the same value in both $YBa_2Cu_3O_{6+x}$ and $Y_{0.8}(Ca)_{0.2}Ba_2Cu_3O_{6+x}$ systems, despite the fact that they are homologous in so many aspects (the same crystal structure, etc.). It should be kept in mind, however, that the arguments raised above are far from being sufficient to completely discard the idea of universal value of the $l_{min}$ parameter in various $Y_{1-b}(Ca)_bBa_2Cu_3O_{6+x}$ systems, so that the issue of possible $b$-dependence of $l_{min}$ remains to be additionally studied. On the other hand, if it is to be spoken in terms of *universal* $l_{min}$, both the results presented here and in reference [26] suggest that it might be $l_{min}$=4.

As regards the χ parameter, it is obvious that introduction of Ca (at Y positions) takes away some more electrons from the planes than in the parent $YBa_2Cu_3O_{6+x}$ system. This additional reduction of electron density in the planes immediately above (below) the chains clearly diminishes chance for a chain hole to capture an electron, and therefore χ($b$) is expected to be descending function, which is exactly what we obtain (χ≈41% ($b$=0), χ≈36% ($b$=0.1), and χ≈33% ($b$=0.2)).

The quantities χ and $l_{min}$ that have been introduced here (as well as in our previous publications [22, 26]) may be further used to characterize some other microscopic aspects that are related to the charge transfer from chains to planes. For example, the estimated value of the χ parameter can be used to evaluate concentration of holes, $n_h$, along the chains (defined as the number of holes that have not been transferred per chains Cu), and to relate it to the *x*-dependent wavelength of charge oscillations, $\lambda_b(x)$, that are stabilized in long chains (for various substitution levels $b$). Although this particular problem will be addressed to in one of our following contributions on wider grounds [29], it is worthwhile to mention that, e. q. for $YBa_2Cu_3O_{6+x}$ system ($b$=0) at $x$≈1, the wave vector $k=2\pi/\lambda$ of charge oscillations is conventionally associated with the $2k_F$ instability (charge density waves along the chains), where the wave vector at the Fermi level is connected to the non-transferred hole concentration along the chains $n_h=1-2p(x≈1)=1-\chi≈0.58(9)$ through the following relation $k_F=n_h\pi/2d$ [30] ($d$≈0.4nm stands for the side of the unit cell parallel to the chains). This gives $\lambda \approx 1.38 nm$ which correlates very well with $\lambda = 1.4 nm$ that has been found in STM experiments [31,32]. In the $Y_{0.8}(Ca)_{0.2}Ba_2Cu_3O_{6+x}$ system the same reasoning points to $\lambda \approx 1.20 nm$ at $x$ close to 1 (unfortunately, no STM data on $\lambda$ have been reported for this system in the literature up to now, at least, as far as we know). In the similar fashion, using the chain length distribution $f(l)$, it is possible to extract $x$- and $b$-dependences of the charge corrugation wavelength, $\lambda_b(x)$, for the whole class of $Y_{1-b}(Ca)_bBa_2Cu_3O_{6+x}$ compounds [29].

## Acknowledgements

This work has been funded by the Ministry of Science and Technology of the Republic of Serbia through the Project 141014.

## Figure captions

**Figure 1.** Calculated $p_{ch}(x)$ and $p(x)=(b_{eff}/2)+p_{ch}(x)$ dependences at $\tau=0.38=const$ are shown by solid lines. For $l_{min}=3$ the two calculated curves are shown: the extracted chain contribution $p_{ch}(x)$, at the bottom part, and $p(x)=(b_{eff}/2)+p_{ch}(x)$ for $b_{eff}=0.154$ and $\chi=0.336$ lying immediately above it (that nearly overlaps with the sequence of small square symbols). In addition, three $p_{ch}(x)$ dependences for $l_{min}=4$, 5, and 6 are also shown, but shifted upwards for 0.12 ($\chi=0.334$), 0.18 ($\chi=0.334$) and 0.24 ($\chi=0.335$), respectively, in order for their behavior to be followed transparently. Open squares denote $p(x)$ values that were derived from experimentally obtained $T_c(x)$ [9] by applying generic $T_c$ versus $p$ relation (1).

**Figure 2.** Calculated values of $T_c(x)$ at $\tau=0.38=const$ are shown by solid line and experimentally obtained values of $T_c(x)$ dependence, as scanned from Figure 4 of reference [9], are shown by open squares, a) for $l_{min}=3$ ($\chi=0.336$, $b_{eff}=0.154$), b) $l_{min}=4$ ($\chi=0.334$, $b_{eff}=0.155$), and c) $l_{min}=5$ ($\chi=0.334$, $b_{eff}=0.156$).

## References

[1] J. L. Tallon, T. Benseman, G. V. M. Williams and J. W. Loram, Physica C 415, 9 (2004).
[2] J. L. Tallon, G. V. M. Williams and J. W. Loram, Physica C 338, 9 (2000).
[3] C. Panagopoulos, J. L. Tallon, B. D. Rainford, T. Xiang, J. R. Cooper and C. A. Scott, Phys. Rev. B 66, 064501 (2002).
[4] J. L. Tallon, C. Bernhard, H. Shaked, R. L. Hitterman and J. D. Jorgensen, Phys. Rev. B 51, 12911 (1995).
[5] G. V. M. Williams, J. L. Tallon, R. Michalak and R. Dupree, Phys. Rev. B 54, 6909 (1996).
[6] R. Liang, D. A. Bonn and W. N. Hardy, Phys. Rev. B 73, 180505 (2006).
[7] J. D. Jorgensen, M. A. Beno, D. G. Hinks, L. Soderholm, K. J. Volin, R. L. Hitterman, J. D. Grace, J. K. Schulle, C. U. Segre, K. Zhang and M. S. Kleefisch, Phys. Rev. B 36, 3608 (1987).

[8] J. L. Tallon and G. V. M. Williams, Phys. Rev. B 61, 9820 (2000).

[9] C. Bernhard and J. L. Tallon, Phys. Rev. B 54, 10201 (1996).

[10] S. Bahrs, A. R. Goffi, C. Thomsen, B. Maiorov, G. Nieva and A. Fainstein, Phys. Rev. B 70, 01451 (2004).

[11] A. Brichhausen, S. Bahrs, K. Fleischer, A. R. Goffi, A. Fainstein, G. Nieva, A. A. Aligia, W. Richter and C. Thomsen, Phys. Rev. B 69, 224508 (2004).

[12] M. Osada, M. Kall, J. Backstrom, M. Kakihana, N. H. Andersen and L. Borjesson, Phys. Rev. 71, 214503 (2005).

[13] S. Bahrs, J. Guimpel, A. R. Goffi, B. Maiorov, A. Fainstein, G. Nieva and C. Thomsen, Phys. Rev. B 72, 144501 (2005).

[14] A. Knizhnik, G. M. Reisner and Y. Eckstein, J. Phys. Cem. Solids 66, 1137 (2005).

[15] H. Shaked, J. D. Jorgensen, B. A. Hunter, R. L. Hitterman, A. P. Paulikas and B. W. Veal, Phys. Rev. B 51, 547 (1995).

[16] A. Sedky, A. Gupta, V. P. S. Awana and A. V. Narlikar, Phys. Rev. B 58, 12495 (1998).

[17] L. T. Wille and D. de Fontaine, Phys. Rev. B 37, 2227 (1988).

[18] V. M. Matic and N. Dj. Lazarov, Physica C 443, 49 (2006).

[19] P. A. Sterne and L. T. Wille, Physica C 162-164, 223 (1989).

[20] D. J. Liu, L. T. Einstein, P. A. Sterne and L. T. Wille, Phys. Rev. B 52, 9784 (1995).

[21] V. M. Matic, Physica A 184, 571 (1992).

[22] V. M. Matic and N. Dj. Lazarov, Solid State Commun. 142, 165 (2007) (arXiv: cond-mat/0611214).

[23] P. Manca, S. Sanna, G. Calestani, A. Migliori, R. De Renzi and G. Allodi, Phys. Rev. B 61, 15450 (2000).

[24] P. Gawiec, D. R. Grempel, A. C. Riiser, H. Haugerud and G. Uimin, Phys. Rev. B 53, 5872 (1996).

[25] P. Gawiec, D. R. Grempel, G. Uimin and J. Zittartz, Phys. Rev. B 53, 5880 (1996).

[26] V. M. Matic and N. Dj. Lazarov, J. Phys.: Condens. Matter 19, 346230 (2007) (arXiv:0704.1887).

[27] K. Mitsen and O. M. Ivanenko, JETP Letters 82, 129 (2005).

[28] M.v. Zimmermann, J. R. Schneider, T. Frelo, N. H. Andersen, J. Madsen, J. Kall, H. F. Poulsen, R. Liang, P. Dosanjih and W. N. Hardy, Phys. Rev. B 68, 104515 (2003).

[29] V. M. Matic and N. Dj. Lazarov, to be submitted.

[30] A. W. Overhauser, Phys. Rev. Lett. 4, 462 (1960);

[31] M. Maki, T. Nishizaki, K. Shibata and N. Kobayashi, Phys. Rev. B 65, 140511 (2002).

[32] D. J. Derro, E. W. Hudson, K. M. Lang, S. H. Pan, J. C. Davis, J. T. Markert and A. L. de Lozanne, Phys. Rev. Lett. 88, 097002 (2002)

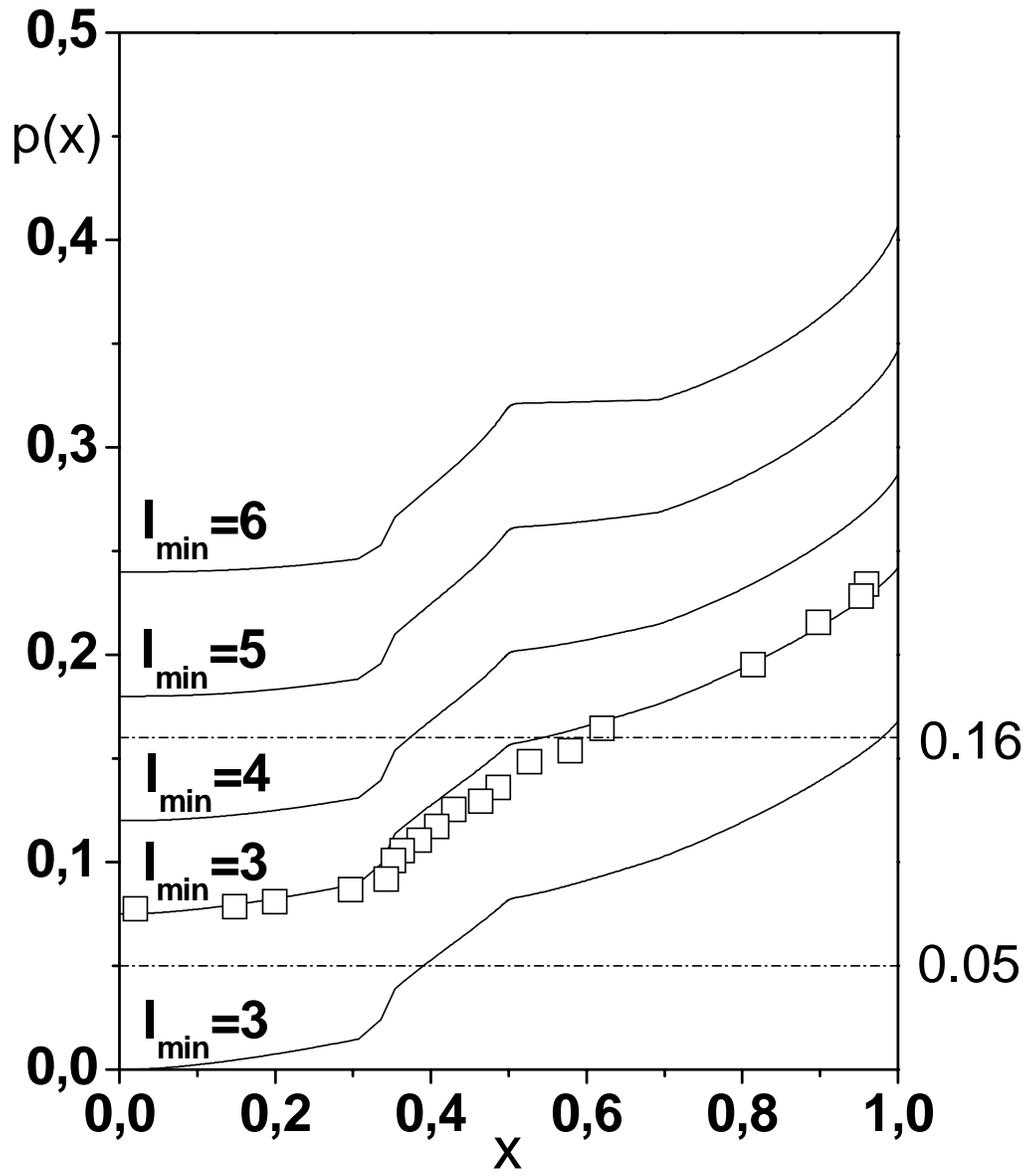

**Figure 1**

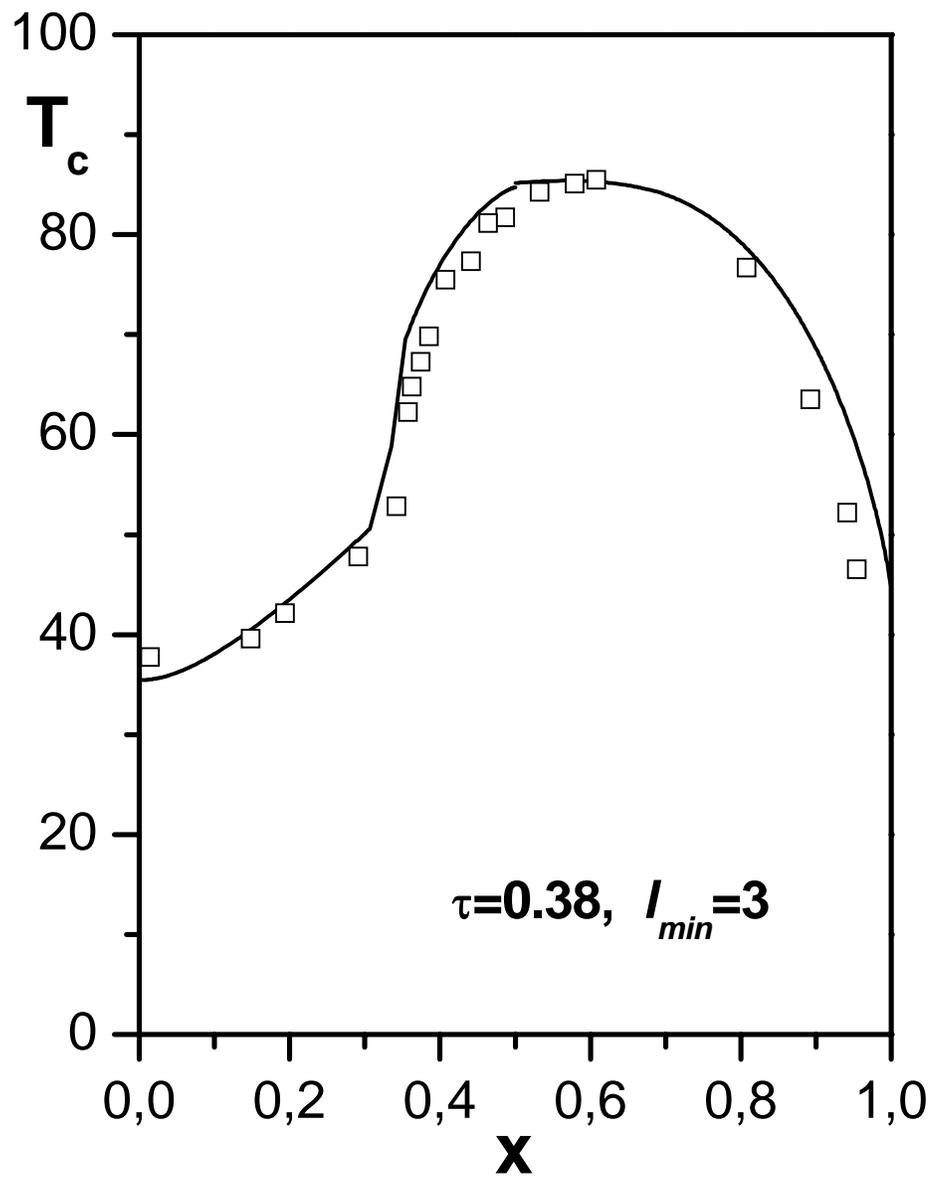

**Figure 2a**

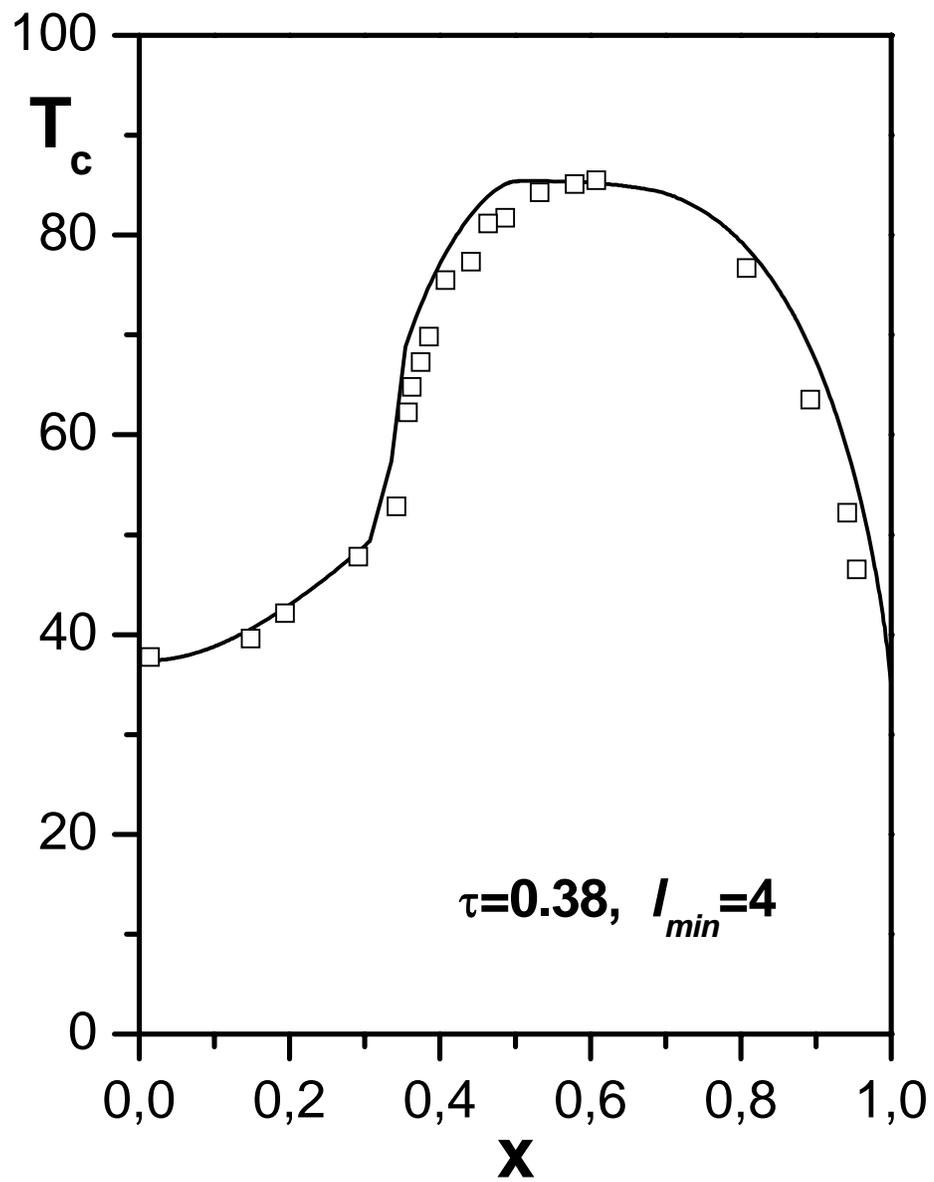

**Figure 2b**

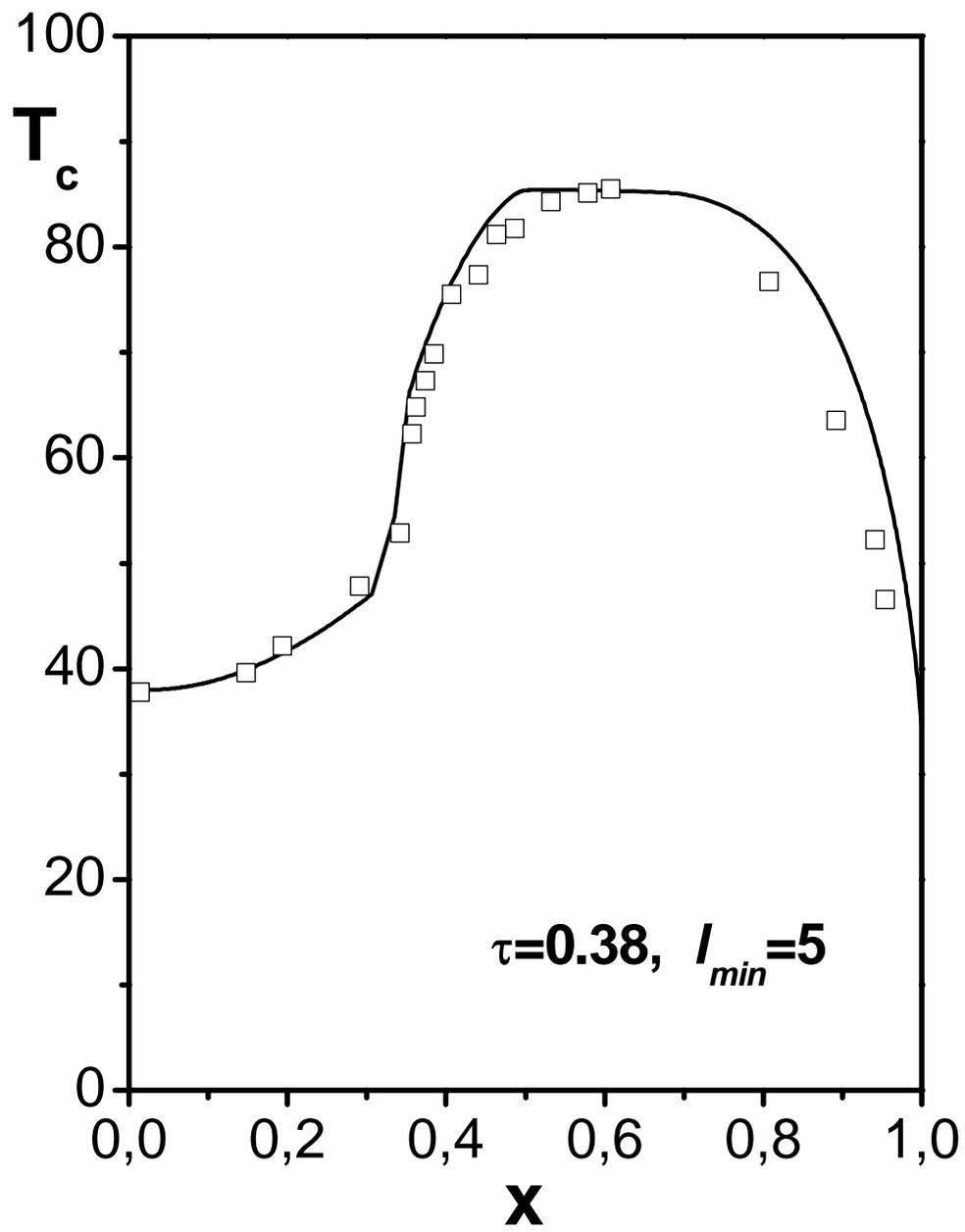

**Figure 2c**